\documentclass[superscriptaddress,prd,11pt]{revtex4-1}
\usepackage{amssymb,amsmath,graphicx,amscd,xcolor,amsthm}
\usepackage{braket}
\usepackage{verbatim} 
\usepackage[colorlinks,citecolor=blue,urlcolor=blue]{hyperref}
\usepackage[caption=false]{subfig}
\usepackage[T1]{fontenc} 

\def\x{\mathbf{x}}
\def\k{\mathbf{k}}

\newcommand{\R}{\mathrm{Re}}
\newcommand{\e}{\mathrm{e}}
\def\d{\mathrm{d}}

\begin{document}

\title{Vacuum induced dispersions on the motion of test particles in D+1 dimensions}
\author{G. H. S. \surname{Camargo}}
\email{guilhermehenrique@unifei.edu.br}
\affiliation{Instituto de Ci\^encias Exatas, Universidade Federal de
Juiz de Fora, Juiz de Fora, Minas Gerais 36036-330,  Brazil}
\author{V. A. \surname{De Lorenci}}
\email{delorenci@unifei.edu.br}
\affiliation{Instituto de F\'{\i}sica e Qu\'{\i}mica, Universidade Federal de Itajub\'a,
Itajub\'a, Minas Gerais 37500-903, Brazil}
\author{C. C. H. \surname{Ribeiro}}
\email{caiocesarribeiro@ifsc.usp.br}
\affiliation{Instituto de F\'{\i}sica de S\~ao Carlos, Universidade de S\~ao Paulo,
S\~ao Carlos, S\~ao Paulo 15980-900, Brazil} 
\author{F. F.  \surname{Rodrigues}}
\email{fernanda-fr@unifei.edu.br}
\affiliation{Instituto de F\'{\i}sica ``Gleb Wataghi'', Universidade Estadual de Campinas,
Campinas, S\~ao Paulo 13083-970,  Brazil} 

\date\today

\begin{abstract}
When the vacuum state of a scalar or electromagnetic field is modified by the presence of a reflecting boundary, an interacting test particle undergoes velocity fluctuations. Such effect is regarded as a sort of quantum analog of the classical Brownian motion.
Several aspects about this system have been recently investigated in the literature, for instance, finite temperature effects, curved spacetime framework, near-boundary regime, late time behavior, and subvacuum phenomena. 
Here, further steps are given in this analysis by considering the effect of vacuum fluctuations of a scalar field in the presence of a perfectly reflecting flat boundary over the motion of a scalar test particle when the background field does not satisfy the Huygens' principle. Specifically, the background field is allowed to have mass and the system is studied in $D+1$ dimensions.
A  method of implementing a smooth transition between distinct states of the field is also developed, rendering regularized analytic expressions describing the velocity fluctuations of the test particle. This method is applied to study some special behaviors of the system. 
Possible applications include fields known to occur in nature as, for instance, the massive Higgs' field, for which the velocity fluctuations are here predicted to acquire a characteristic oscillation, thus behaving differently from their electromagnetic counterparts. 
\end{abstract}


\maketitle

\section{Introduction}
\label{intro}
Transition between vacuum states of the background electromagnetic field induces velocity fluctuations over a charged test particle that is interacting with this field. This phenomenon was originally reported \cite{ford2004} for a nonrelativistic particle released initially at rest nearby a perfectly conducting flat boundary, and since then, it has been explored in various possible ways. For instance, finite temperature effects were included \cite{hongwei2006}, and the idea was applied to study velocity fluctuations in Robertson-Walker spacetimes, where the varying scale factor is responsible for driving the vacuum transition \cite{bessa2009}. Among the features unveiled by this effect is the change of the kinetic energy of the particle during the transition between vacuum states. In the case of an electric charged particle, when the vacuum state is modified by the presence of a perfectly reflecting boundary placed at $z=0$, it was shown that in the late time regime the particle kinetic energy per unit mass is increased by the amount $\alpha(2\pi z^2)^{-1}$, where $\alpha$ is the fine-structure constant. The observed divergence at the wall was linked to the use of idealized boundary conditions. Another divergence, also connected to the idealization of the model, was reported to occur in a time $t=2z$ after the particle is released in the presence of the modified vacuum.

A model including quantum aspects of the particle was also examined \cite{Seriu2009} as an attempt to bring more reality to the system, and it was shown that such procedure is enough to regularize the late time regime. However, the model was not completely integrated, and only particular expansions were obtained. In particular, the behavior at $t=2z$ could not be addressed. The use of a smooth switching to study the late time regime of this system was also examined \cite{seriu2008}. A technique of distance fluctuations to study the simplified $1+1$ dimensional scalar model  was successfully implemented \cite{delorenci2014}, and despite its regularization at the wall and at $t=2z$, higher dimensional cases were not considered, and a new kind of divergence was reported to occur at late times. 
Recently a more complete description based on smooth switching techniques to model the vacuum transition was reported \cite{delorenci2016,delorenci2019b}. It was shown that the smooth character of the transition is enough to regularize all the velocity dispersions, without changing the idealized boundary hypothesis. This idea was further developed to include a general switching in the case of $3+1$ scalar field \cite{Camargo2018}, where new features were unveiled. For instance, depending on the measurement setup, the particle kinetic energy can be lessened by a certain amount, a sort of subvacuum phenomena, where strictly positive quantities at classical level become negative at a quantum level \cite{delorenci2019}.

It should be stressed that the idea of using switching mechanisms is close to the one of averaging observables. Just to establish the connection, a perfectly conducting boundary, as the one used to study velocity fluctuations, polarizes the vacuum around it giving rise to fluctuating Casimir forces that diverge
 at the boundary. A possible way of bringing reality to this physically unfeasible model is by implementing averaged stresses, that regularize the divergences and enables one to study the fluctuations near the boundary \cite{Barton1991A,Barton1991B}. In fact, actual measurements of these stresses are only meaningful for time and space averages. Another instance where this technique was implemented to bring physical meaning to the observables was in a recently investigated analog model for light cone fluctuations due to stress tensor fluctuations \cite{bessa2016}, where the geometry of a nonlinear dielectric slab naturally provided a space averaged quantum observable, thus regularizing the model.     

In this work, we go a step further in  the analysis of the velocity fluctuations of a test particle in the presence of a perfectly reflecting flat boundary when the background scalar field is allowed to have a nonvanishing mass. The physical motivation for studying this system relies on the fact that only massive scalar fields are known to occur in nature, e.g., the Higgs' field. We also let the spacetime dimension to be arbitrary, thus including previous studies as particular cases. 
One important motivation in going through this generalization is the possibility to examine frameworks where the background field fails to satisfy the Huygens' principle. 
In such circumstances, the fluctuations significantly depart from previously known cases. In order to integrate the velocity dispersions, we present a general discussion of switching functions in frequency space, revealing the physical mechanism behind the regularization process, and how to model particular switchings that agree with the system symmetry. 

The paper is organized as follows. The next section establishes the basic Langevin equation modeling the effect. In Sec.~\ref{secIII} the quantization of the massive scalar field in D+1 spacetime dimensions is presented, and Huygens' principle is defined. In Sec.~\ref{secIV}, we study switching mechanisms in frequency space, unveiling some general regularization aspects. We present the velocities fluctuations in Sec.~\ref{secV}, and close with final remarks in Sec.~\ref{secVI}. The Appendix contains a detailed calculation of the late time regime. Units are such that $c=\hbar=1$. 

\section{Langevin equation}
\label{secII}
The system under study consists of a massive real scalar field $\phi(t,\x)$ in $D+1$ dimensions, with $D$ a positive integer number, and a non-relativistic point-like scalar particle of charge to mass ratio $g$. Here, the vector $\x$ is written as $(x_1,\ldots,x_D)$. In the nonrelativistic regime, the particle dynamics is governed by the Newtonian force law \cite{Camargo2018}
\begin{equation}
\frac{dv_i}{dt}=-g\frac{\partial\phi}{\partial x_i},\label{eq1}
\end{equation}
where $x_i=x_i(t)$ denotes the particle position, and $v_i=dx_i/dt=\dot{x}_i$ its velocity. The subscript index $i$ runs from $1$ to $D$, denoting the $i$th Cartesian component. This particle is assumed to be a test particle in the sense the field it creates is small, and can be neglected from the analysis. 
%

The scalar particle works as a probe for quantum fluctuations of the background field when it is prepared in some vacuum state, for which $\braket{\phi(t,\x)}\equiv\bra{0}\phi(t,\x)\ket{0}=0$. Hence, Eq.~\eqref{eq1}  becomes a Langevin-like equation for the particle position, and its acceleration vanishes on average, $\braket{\dot{v}_i}=0$. Despite this, it will in general fluctuates around this average, as negative fluctuations, when squared, do not cancel in the averaging process, and thus $\braket{\dot{v}_i^2}\neq0$. If we assume the particle displacement to be negligible during the interaction with the field, its final velocity will be
\begin{equation}
v_i(\tau)=-g\frac{\partial\ }{\partial x_i}\int_0^{\tau}\d t'\phi(t',\x),\label{eq2}
\end{equation}
where $\tau$ is hereafter called the measuring time. Thus, a measurement of the velocity component $v_i$ will be distributed around its average, $\braket{v_i}=0$, with mean square deviation $\braket{(\Delta v_i)^2}=\braket{v_i^2}$ due to quantum vacuum fluctuations.

We will assume the field to be in its modified vacuum state due to the presence of a perfectly reflecting plane boundary. Recently, this scenario was studied for a massless scalar field in $3+1$ dimensions \cite{Camargo2018}. Let us recall some of its features to set the notation up. If we call $d$ the distance from the particle to the wall, it was shown that when the interaction is instantaneously turned on at $t=0s$, and turned off at $t=\tau>0$, the dispersions $\braket{(\Delta v_i)^2}$ diverge at the wall and also when $\tau=2d$, which corresponds to the time of a signal's round trip between particle and boundary. It was shown that adding the smooth character of the interaction to the model is enough to regularize these divergences. The physical mechanism behind it is the finite amount of time, hereafter called the switching time $\tau_s>0$, needed for the particle to be released in its initial position, and also for completing the measurement afterwards. It is also noteworthy that the particle is supposed not to disturb the background field. Therefore, the smooth switching is equivalent to a time dependent coupling $gF(t)$, where the switching function $F(t)$ is normalized according to
\begin{equation}
\int_{-\infty}^{\infty}\d t F(t)=\tau.\label{eq3}
\end{equation}        
Thus the velocity dispersions become
\begin{equation}
\braket{(\Delta v_i)^2}=\frac{g^2}{2}\left[\frac{\partial\ }{\partial x_i}\frac{\partial\ }{\partial x_i'}\int_{-\infty}^{\infty}\d tF(t)\int_{-\infty}^{\infty}\d t'F(t')G^{(1)}(t,\x;t',\x')\right]_{\x'=\x},\label{eq4}
\end{equation}
with $G^{(1)}(t,\x;t',\x')=\braket{\phi(t,\x)\phi(t',\x')+\phi(t',\x')\phi(t,\x)}$. Notice that the sudden switching process corresponds to set $F(t)=\Theta(t)\Theta(\tau-t)$, $\Theta(t)$ being the unit step function. The function $F(t)$ thus defined is a dynamical quantity that models how the experiment is done. We will explore some of its physical properties later on. 

\section{Aspects of the quantized field}
\label{secIII}
This section reviews the canonical quantization of the massive scalar field in the presence of a perfectly reflecting plane wall in $D+1$ dimensions. The wall is placed at $x_1=0$ in all cases. We call especial attention to the propagation of signals and nonlocal effects when the field has non-vanishing mass, and in the massless case when $D$ is an even number. As we will see, these cases present features contrasting significantly from the massless $3$-dimensional case.

We start by studying the quantization of the free field. Generalization to the case where a boundary is present can easily be implemented by using the method of images. Let $m$ be the field mass. Its dynamics is given by the free Klein-Gordon equation $(\Box+m^2)\phi=0$, where $\Box=\eta^{\mu\nu}\partial_\mu\partial_\nu$, and $\eta^{\mu\nu}$ is the Minkowski metric in Cartesian coordinates. The quantized field is written in terms of the creation/annihilation operators as $\phi(t,\x)=\int\d^Dk\left[a_{\k}u_{\k}(t,\x)+a^\dagger_{\k}u^*_{\k}(t,\x)\right]$, where $\left[a_{\k},a^\dagger_{\k'}\right]=\delta^D(\k-\k')$, and all the other commutators vanish. The properly normalized eigenfunctions $u_\k$ are \cite{Davies1982}
\begin{equation}
u_{\k}(t,\x)=\frac{1}{\sqrt{2\omega(2\pi)^D}}\e^{-i\omega t+i\k\cdot\x},
\label{eq5}
\end{equation}
where the wave vector $\k \in \mathbb{R}^D$, and $\omega=\sqrt{\k^2+m^2}$. The vacuum state $\ket{0}$ is such that $a_{\k}\ket{0}=0$ for all $\k$. The propagators for the real scalar field can be written in terms of the two-point (Wightman) function $G^{+}(t,\x;t',\x')$, defined as the vacuum expectation $G^{+}(t,\x;t',\x')=\braket{\phi(t,\x)\phi(t',\x')}$. It can be shown that, after using a generalized spherical coordinate transformation, this function has the following integral representation in terms of Bessel functions 
\begin{equation}
G^{+}(t,\x;t',\x')=\frac{1}{2(2\pi)^\frac{D}{2}|\Delta\x|^{\frac{D}{2}-1}}\int_0^\infty\d k\frac{k}{\omega}\e^{-i\omega\Delta t}k^{\frac{D}{2}-1}J_{\frac{D}{2}-1}(k|\Delta\x|),\label{eq6}
\end{equation}
and can be exactly integrated as \cite{gradshteyn}
\begin{equation}
G^{+}(t,\x;t',\x')=\frac{1}{2\pi}\left(\frac{m}{2\pi i \sigma}\right)^{\frac{D-1}{2}}K_{\frac{D-1}{2}}(im\sigma).
\label{eq7}
\end{equation}
Here, $K_\nu(z)$ is the modified Bessel function of the second kind, $\sigma=\sqrt{(\Delta t)^2-(\Delta\x)^2}$, and $\Delta t=t-t'-i\epsilon$, with $\epsilon$ a small positive real number added to ensure convergence. In what follows, we will use two propagators, the Hadamard two-point function $G^{(1)}(t,\x;t',\x')=2\mbox{Re}\ G^{+}(t,\x;t',\x')$ appearing in Eq.~\eqref{eq4}, and the retarded scalar propagator, defined as $G_{R}(t,\x;t',\x')=-2\Theta(t-t')\ \mbox{Im}\ G^{+}(t,\x;t',\x')$.
Thus the canonical commutation relations imply that $(\Box+m^2)G_{R}(t,\x;t',\x')=\delta(t-t')\delta^D(\x-\x')$, that is, the retarded propagator is the field intensity at $(t,\x)$ produced by a deltalike unit charge at $(t',\x')$. It can be shown that the support of $G_{R}(t,\x;t',\x')$ is contained in the future emission of $(t',\x')$, i.e., the set $J^{+}(t',\x')=\{(t,\x):t>t'\ \mbox{and}\ (\Delta t)^2\ge(\Delta\x)^2\}$ \cite{Friedlander1975}.

As anticipated, the present work extends previous analysis \cite{Camargo2018} by considering the effect of field mass and spacetime dimension on vacuum induced velocity fluctuations of scalar particles. In contrast to previous works, now the background field is such that its wave equation does not satisfy the Huygens' principle, defined as follows \cite{Friedlander1975}. The wave equation $(\Box+m^2)\phi=0$ is said to satisfy this principle if the corresponding retarded propagator $G_{R}(t,\x;t',\x')$ is supported on the future light cone $C^{+}(t',\x')=\{(t,\x):t>t'\ \mbox{and}\ (\Delta t)^2=(\Delta\x)^2\}$. In physical terms, if a wave equation is of Huygens type, then field disturbances that have a sudden beginning also have a sudden ending. 

\begin{figure}[h!]
\center
\includegraphics[scale=0.55]{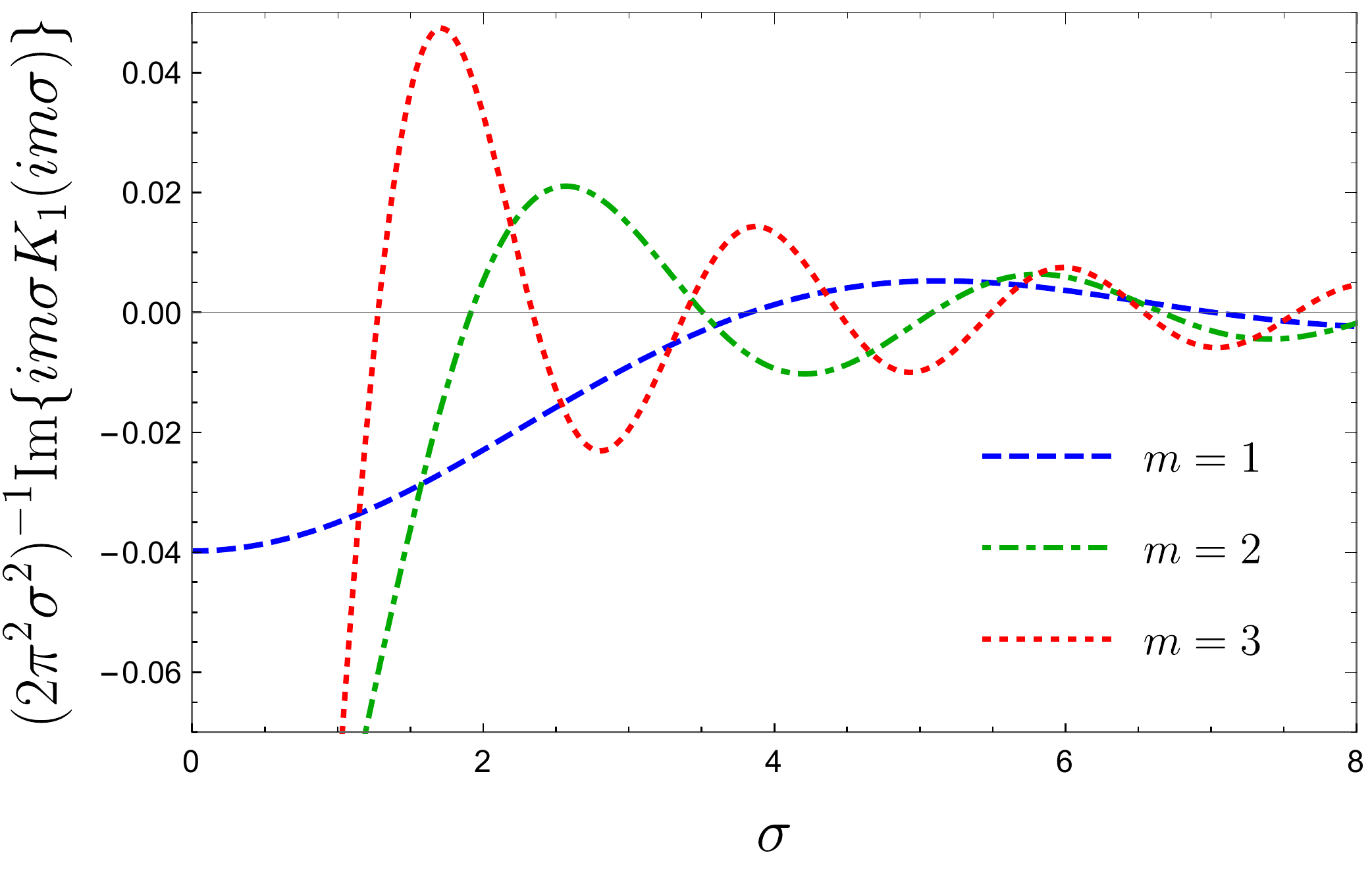}
\caption{Some representative plots for the propagator tail as function of $\sigma$. Here, $D=3$.}
\label{fig1}
\end{figure}

Let us analyze the massive $D=3$ case, for which Eq.~\eqref{eq7} reduces to
\begin{equation}
G_{R}(t,\x;t',\x')=\frac{\Theta(t-t')}{2\pi}\delta(\sigma^2)+\frac{\Theta(t-t')}{2\pi^2\sigma^2}\mbox{Im}\!\left\{ im\sigma K_{1}(im\sigma)\right\},
\label{eq9}
\end{equation}
where the limit $\epsilon\rightarrow 0$ was taken. 
 As we see, the propagator is composed by a sum of a term supported on the future light cone $\sigma^2=0$ plus a ``tail'' term that does not vanish inside the light cone, $\sigma^2>0$, if $m>0$. Thus an inertial observer will experience the field produced in an early time by a deltalike source as a sudden impulse followed by a never ceasing retarded signal. The particular form of the tail term depends on how the field internal degrees of freedom individually propagate signals. In this case, their net effect presents nonintuitive interference patterns, as depicted in Fig.~\ref{fig1}. 
It is possible to track the origin of the non-Huygensian character down by looking at the individual plane wave solutions given by Eq.~\eqref{eq5}. Notice that these waves present dispersion, and their group velocity is $v_{g}=k/\sqrt{k^2+m^2}$. Thus, the field possesses modes carrying information at arbitrarily low velocities, resulting, for instance, in the interference patterns of Fig.~\ref{fig1}.

\begin{figure}[h!]
\center
\includegraphics[scale=0.55]{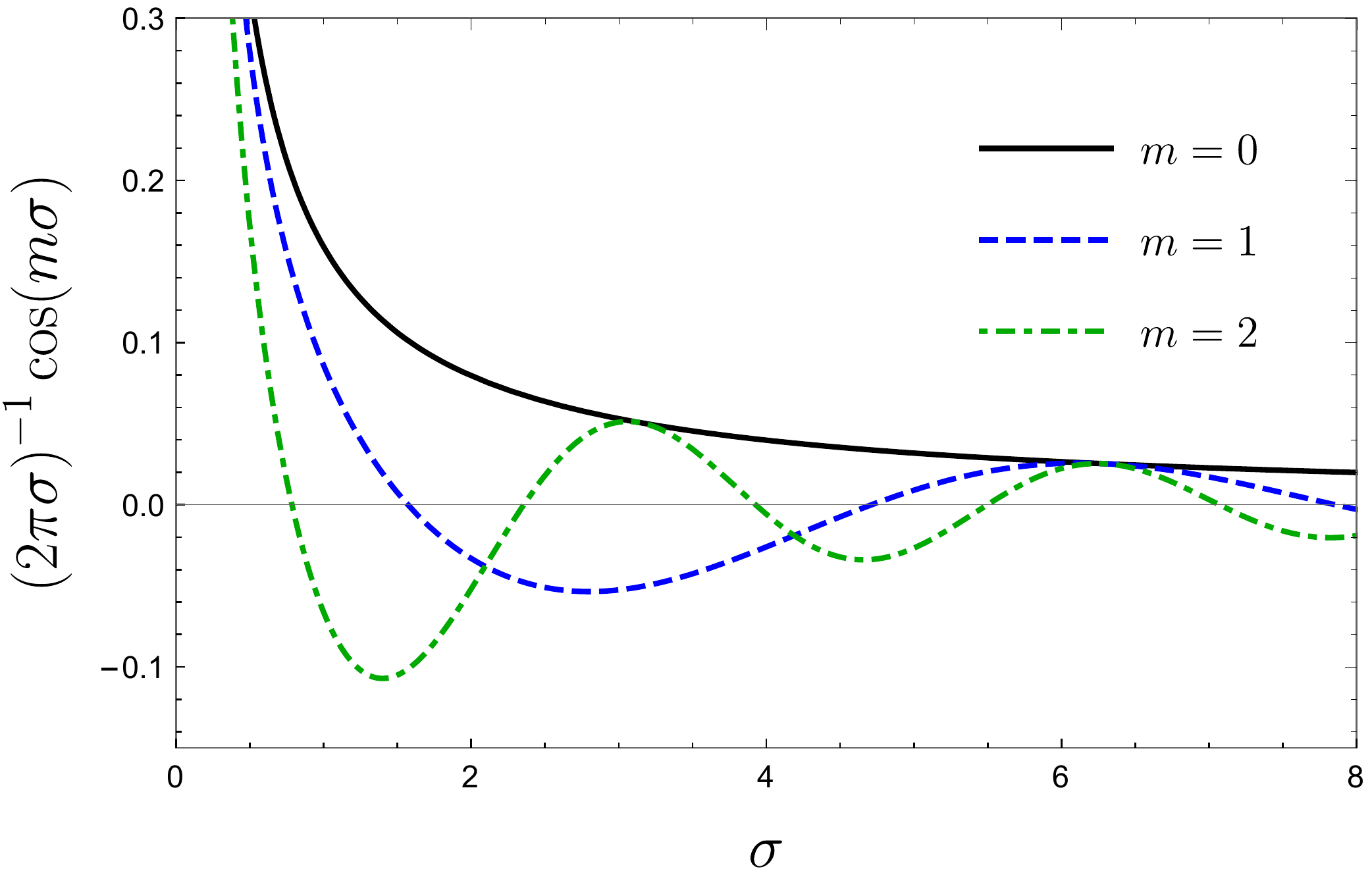}
\caption{Some representative plots for the propagator tail as function of $\sigma$ when $D=2$. In this case the plot corresponding to the massless field works as an envelope providing a maximum for the oscillation amplitude.}
\label{fig2}
\end{figure}

The case $D=2$ is richer, for even the massless case is non-Huygensian, as it is shown in Fig.~\ref{fig2}. In fact, Huygens' principle does not hold for the d'Alembertian in odd dimensional flat spacetimes \cite{Friedlander1975}.
Therefore, in contrast to the former case, the non-Huygensian character of $D=2$ does not possess a simple interpretation in terms of plane waves.

As for the $D=1$ case, inspection of Eq.~\eqref{eq6} or \eqref{eq7} reveals that the propagator is not defined for the massless case. In fact, the observed infrared behavior is well known in quantum field theory. Nevertheless, the dispersion can be calculated in the $1+1$ case taking the field to have a small mass and making it go to zero at the end of the calculation, as was done in \cite{delorenci2014}. Thus, this case fails to satisfy the Huygens' principle by the same reason as the $3+1$ case.

The propagation of field disturbances is an example of nonlocal effect, and the above discussion indicates how it is related to the Huygens' principle. The vacuum induced velocity dispersions $\braket{(\Delta v_i)^2}$, defined through Eq.~\eqref{eq4}, depend on field averages over an extended period of time, and thus are also nonlocal effects. We will see in Sec.~\ref{secV} that the dispersions present similar features coming from the non-Huygensian character of the background field.

We close this section with the quantized field in the presence of a reflecting flat wall placed at $x_1=0$. By the method of images, it is straightforward to see that the Wightman function is amended as
\begin{equation}
G^{+}(t,\x;t',\x')=G^{+}_{\tt 0}(t,\x;t',\x')-\frac{1}{2\pi}\left(\frac{m}{2\pi i \sigma_{+}}\right)^{\frac{D-1}{2}}K_{\frac{D-1}{2}}(im\sigma_{+}),
\label{eq10}
\end{equation}
where $\sigma_{+}=\sqrt{(\Delta t)^2-(\hat{\Delta} \x)^2}$, and $\hat{\Delta} \x$ is the vector obtained from $\Delta \x$ by exchanging $x_1'\rightarrow-x_1'$. In the above equation $G^{+}_{\tt 0}(t,\x;t',\x')$ is the Minkowskian two-point function defined in Eq.~\eqref{eq7}. By linearity, all other propagators are readily obtained in terms of the modified Wightman function. In particular, the Hadamard two-point function is  $G^{(1)}(t,\x;t',\x')=G^{(1)}_{\tt 0}(t,\x;t',\x')+G^{(1)}_{\tt Ren}(t,\x;t',\x')$, with
\begin{equation}
G^{(1)}_{\tt Ren}(t,\x;t',\x')=-\frac{1}{\pi}\mbox{Re}\left\{\left(\frac{m}{2\pi i \sigma_{+}}\right)^{\frac{D-1}{2}}K_{\frac{D-1}{2}}(im\sigma_{+})\right\},\label{eq11}
\end{equation}
where the subscript ${\tt Ren}$ stands for renormalized. In calculating the dispersions $\braket{(\Delta v_i)^2}$, we should keep only the renormalized propagator given by Eq.~\eqref{eq11}, as the Minkowski vacuum do not cause velocity fluctuations on test particles \cite{johnson2002}. 
 
\section{Smooth switching in frequency space}
\label{secIV}
If the renormalized propagator given by Eq.~\eqref{eq11} is written in its integral representation, as it was done with the Wightman function in Eq.~\eqref{eq6}, and inserted in Eq.~\eqref{eq4}, we will obtain that 
\begin{equation}
\braket{(\Delta v_i)^2}=\frac{g^2}{2}\left[\frac{\partial\ }{\partial x_i}\frac{\partial\ }{\partial x_i'}\frac{-1}{(2\pi)^\frac{D}{2}|\hat{\Delta}\x|^{\frac{D}{2}-1}}\int_0^\infty\d k\frac{k}{\omega}|\widehat{F}(\omega)|^2k^{\frac{D}{2}-1}J_{\frac{D}{2}-1}(k|\hat{\Delta}\x|)\right]_{\x'=\x},\label{eq12}
\end{equation}
where we have defined the Fourier transform of $F(t)$ \cite{fewster2015},
\begin{equation}
\widehat{F}(\omega)=\int_{-\infty}^{\infty}\d t\e^{-i\omega t}F(t).\label{eq13}
\end{equation}
The normalization condition expressed by Eq.~\eqref{eq3} is simply the constraint $\widehat{F}(0)=\tau$. Thus, the question of whether or not a given absolute integrable profile $\widehat{F}(\omega)$ satisfying $\widehat{F}(0)=\tau$ corresponds to a physically admissible switching naturally arises. For instance, a necessary condition is $\widehat{F}(\omega)=\widehat{F}^*(-\omega)$ for $F(t)$ to be real. This question can be partially answered once we establish what is meant by an admissible switching. By its very construction, we should assume that such $F(t)$ do not alter the ``sign'' of the interaction, and is limited in magnitude by the maximum possible value, $1$. Thus, a switching is admissible if $0\leq F(t)\leq1$. In this case, $F(t)/\tau$ can be mathematically viewed as a probability density function, and the Bochner theorem \cite{Rudin1990} states that $F(t)$ is a probability density function if, and only if, its Fourier transform $\widehat{F}(\omega)$ is continuous, normalized as $\widehat{F}(0)=\tau$, and for every positive integer number $n$, the positivity condition
\begin{equation}
\sum_{i,j=1}^{n}\widehat{F}(\omega_i-\omega_j)\xi_{i}\xi_{j}^*\geq0\label{eq13-2}
\end{equation}
holds for arbitrary real numbers $\omega_{i}$ and complex numbers $\xi_{i}$. In practice, though, Eq.~\eqref{eq13-2} is not enlightening. Nevertheless, it can be used to guide one's search for a switching in frequency space. In fact, for $n=1,2$ this condition simply means that $\widehat{F}(0)\geq 0$ and $|\widehat{F}(\omega)|\leq \widehat{F}(0)$. Moreover, given a Fourier transform such that the corresponding function $F(t)$ is a probability distribution, one must ensure that $F(t)$ is bounded by $1$ in order to be an admissible switching. We will use these remarks to motivate the use of a particular $F(t)$ later on.

A convenient choice of the switching function is $F^{(1)}_{n,\tau}(t)=c_{n}/[1+(2t/\tau)^{2n}]$, with $c_{n}=(2n/\pi)\sin (\pi/2n)$ \cite{delorenci2016}. Let us quote some of its properties. As $n\rightarrow\infty$, it is clear that $F^{(1)}_{n,\tau}(t)$ recovers the sudden switching mechanism. Thus, for a finite $n$, it models a smooth extension of the former. The switching time $\tau_s$ is fixed by the measurement duration $\tau$ and the parameter $n$, and for large $n$ it is approximately given by $\tau_s=(\tau/2n)\ln(2+\sqrt{3})$. Therefore, for a fixed large $n$, $\tau_s$ grows linearly with $\tau$, and there will be no residual effects in the late time regime $\tau\rightarrow\infty$. Despite this loss of information in the late time regime, the switching $F^{(1)}_{n,\tau}(t)$ is well suitable to study the dispersions for a finite $\tau$, and the integral in Eq.~\eqref{eq12} can be performed using the Fourier transform of $F^{(1)}_{n,\tau}(t)$, which in turn is found from the results in \cite{delorenci2016} to be
\begin{equation}
\widehat{F}^{(1)}_{n,\tau}(\omega)=\frac{i\tau\pi c_n}{2n}\sum_{q=n}^{2n-1}\psi_{n,q}\e^{-i\omega\tau\psi_{n,q}/2},\label{eq14}
\end{equation}
with $\psi_{n,p}=\exp[i(\pi/2n)(1+2p)]$.%

In order to gain control of the late time regime, a switching with controllable $\tau_s$ must be implemented. For instance, the choice
\begin{equation}
F_{\tau_s,\tau}^{(2)}(t)=\frac{1}{\pi}\left[\arctan \left(\frac{t}{\tau_s}\right)+\arctan \left(\frac{\tau-t}{\tau_s}\right)\right]\label{eq15}
\end{equation} 
was successfully used to analyze the dispersions in the massless $3+1$ scalar case for all $\tau$ \cite{Camargo2018}, and more recently it was implemented to study the near-boundary regime for the electromagnetic case \cite{delorenci2019b}. Nevertheless, dispersions modeled by this switching have the drawback of depending on complex arguments, which are not of easy manipulation. Moreover, when $m\ne 0$ the integral in Eq.~\eqref{eq12} does not seem to have a simple analytic closed expression with the choice of the switching defined by Eq.~\eqref{eq15}. In fact, even for the sudden switching mechanism $F^{(0)}_{\tau}(t)=\Theta(t)\Theta(\tau-t)$, the massive case is not simple.

\begin{figure}[h!]
\center
\includegraphics[scale=0.55]{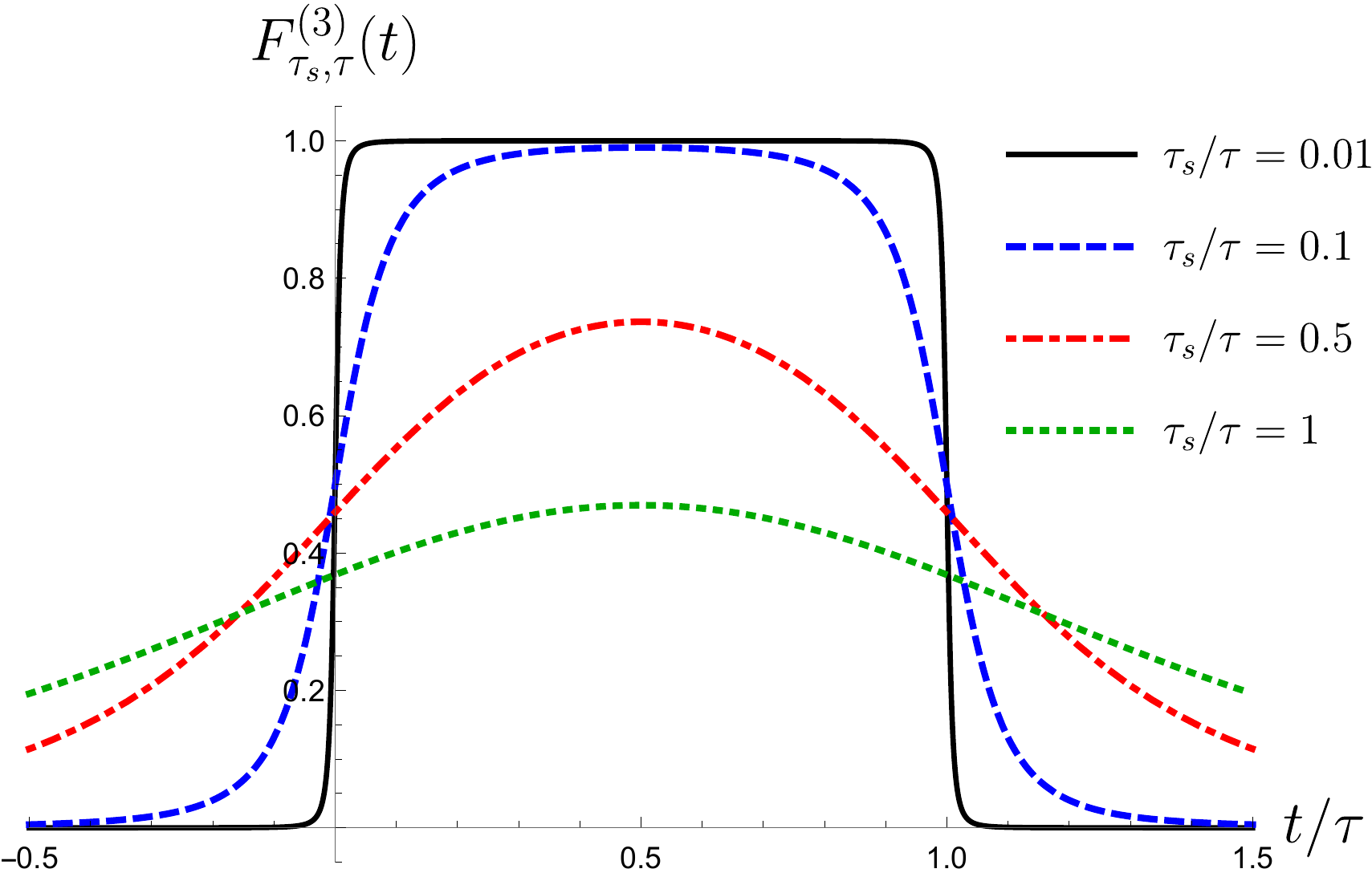}
\caption{Smooth switching behavior of $F^{(3)}_{\tau_s,\tau}(t)$. The sudden limit $F^{(0)}_\tau(t)$ is recovered when $\tau_s\rightarrow0$, as suggested by the plots, and verified from the definition of $F^{(3)}_{\tau_s,\tau}(t)$.}
\label{fig3}
\end{figure}

Despite this, it is possible to construct a suitable switching mechanism with controllable $\tau_s$ for which the late time regime of Eq.~\eqref{eq12} can be explicitly found. In order to do that, we start from the Fourier transform of $F^{(0)}_{\tau}(t)$, $\widehat{F}^{(0)}_{\tau}(\omega)=(1/i\omega)(1-\e^{-i\omega\tau})$. Thus, we will look for a switching function given as the inverse Fourier transform of
\begin{equation}
\widehat{F}(\omega)=\frac{1}{i\omega}(1-\e^{-i\omega\tau})\mathcal{D}(\omega).\label{eq17}
\end{equation}
The normalization $\widehat{F}(0)=\tau$ requires that $\mathcal{D}(0)=1$, and $\mathcal{D}$ should be bounded as $|\omega|\rightarrow\infty$ to ensure convergence. For instance, if $\mathcal{D}(\omega)=\e^{-\tau_s|\omega|}$, the corresponding switching is the one defined by Eq.~\eqref{eq15}. This procedure reveals how the smooth switching regularizes the usual divergences presented by this sort of system. It acts as a filter for high frequency modes. The decaying factor $\mathcal{D}$ must be chosen according to the remarks presented in the beginning of the section. For example, if we set $\mathcal{D}(\omega)=\Theta(1/\tau_s-|\omega|)$, the function $F(t)$ obtained is smooth and tends to $F^{(0)}_{\tau}$ as $\tau_s\rightarrow0$. Nevertheless, the hypotheses of Bochner's theorem are not verified, as $\mathcal{D}$ is not continuous, and thus $F(t)<0$ for some values of $t$. In what follows, we work with the function defined by $\mathcal{D}(\omega)=\sqrt{1+2\tau_s|\omega|}\e^{-\tau_s|\omega|}$, and thus
\begin{equation}
F^{(3)}_{\tau_s,\tau}(t)=\frac{1}{2\pi}\int_{-\infty}^{\infty}\d\omega \frac{\e^{i\omega t}}{i\omega}(1-\e^{-i\omega \tau})\sqrt{1+2\tau_s|\omega|}\e^{-\tau_s|\omega|}.\label{eq18}
\end{equation}
Illustrative profiles modeled by $F^{(3)}_{\tau_s,\tau}(t)$ for some chosen values of $\tau_s$ are depicted in Fig.~\ref{fig3}. The plots were done using a closed form for Eq.~\eqref{eq18} in terms of special functions, from which we verify the condition $0\leq F^{(3)}_{\tau_s,\tau}(t)\leq1$. The full expression, though, is not very clarifying, and thus will be omitted here, as only its Fourier representation $\widehat{F}(\omega)$ [Eq.~\eqref{eq17}] is needed in calculating the dispersions.

\section{Velocity dispersions}
\label{secV}
We are now able to present the various dispersions contemplated by Eq.~\eqref{eq12}. Let us start with a closed expression obtained with the smooth switching $F^{(1)}_{n,\tau}$. Let $\braket{(\Delta v_{x_1})^2}\doteq\braket{(\Delta v_{\bot})^2}^{(1)}_{D}$, and $\braket{(\Delta v_{x_i})^2}\doteq\braket{(\Delta v_{\|})^2}^{(1)}_{D}$, $i>1$. The superscript $(1)$ indicates the measurement done according to $F^{(1)}_{n,\tau}$. Thus, by combining Eqs.~\eqref{eq12} and \eqref{eq14}, and using Eq.~\eqref{eq7}, it follows
\begin{align}
\braket{(\Delta v_{\|})^2}^{(1)}_{D}&=-\frac{g^2}{x^{D-1}}\left[\frac{(\tau/x)\pi c_n}{2n}\right]^2\sum_{p,q=0}^{n-1}\psi_{n,p}\psi_{n,q}^{*}\left(\frac{mx}{4\pi\gamma_{p,q}}\right)^{\frac{D+1}{2}}K_{\frac{D+1}{2}}(2mx\gamma_{p,q}),\label{eq19}\\
\braket{(\Delta v_{\bot})^2}^{(1)}_{D}&=8\pi x^2\braket{(\Delta v_{\|})^2}^{(1)}_{D+2}-\braket{(\Delta v_{\|})^2}^{(1)}_{D},\label{eq20}
\end{align}
where now $\gamma_{p,q}=\left[1-(\tau/4x)^2(\psi_{n,p}-\psi_{n,q}^{*})^2\right]^{\frac{1}{2}}$, and $x=x_{1}$ is the particle distance to the wall. 
The generalization of Eqs.~\eqref{eq19} and \eqref{eq20} is twofold, as the background field is allowed to have any mass $m$ and the dimension $D$ can take any positive integer values. For instance, the results in Ref.~\cite{Camargo2018} can be recovered by setting $D=3$ and take $m\rightarrow0$. Bellow we analyze some of the properties of the results presented in Eqs.~\eqref{eq19} and \eqref{eq20}.

The massless $D=1$ case using the sudden switching, in which only the velocity component perpendicular to the wall is present, was studied in Ref.~\cite{delorenci2014}. The expected divergences at $x=0$ and at $\tau=2x$ were reported, and a mechanism of distance fluctuation was implemented in order to bring more reality to the system. It was shown that this fluctuation is enough to regularize the divergences. Despite this, higher dimensional systems were not addressed, and it is not clear if a fluctuating distance also regularizes them. Moreover, another kind of divergence appears in this system, as the dispersion grows indefinitely as $\tau$ goes to infinity. For $m>0$, by noticing that $\R\ \gamma_{p,q}\rightarrow \infty$ as $\tau\rightarrow\infty$, and using the asymptotic form of the Bessel functions $K_{\nu}(z)\approx \mathrm{e}^{-z}\sqrt{\pi/2z}$ for large $z$ \cite{gradshteyn}, we see that the dispersions in Eq.~\eqref{eq20} are exponentially suppressed in the late time regime, in sharp distinction with the sudden switched massless case. If $m=0$, then the late time regime of Eq.~\eqref{eq20} is $\braket{(\Delta v_{\bot})^2}^{(1)}_{1}=(g^2\pi c_{n}^2/2n^2)\sum_{p,q}\psi_{n,p}\psi_{n,p}^*/(\psi_{n,p}-\psi_{n,q}^*)^2$, a finite constant value. However, we saw in the previous section that the switching mechanism $F^{(1)}_{n,\tau}$ is not well suited for studying the late time behavior, as the switching time becomes infinite. We will return to this matter later, when we present the dispersions calculated via $F^{(3)}_{\tau_s,\tau}(t)$. 

\begin{figure}[h!]
\center
\includegraphics[scale=0.55]{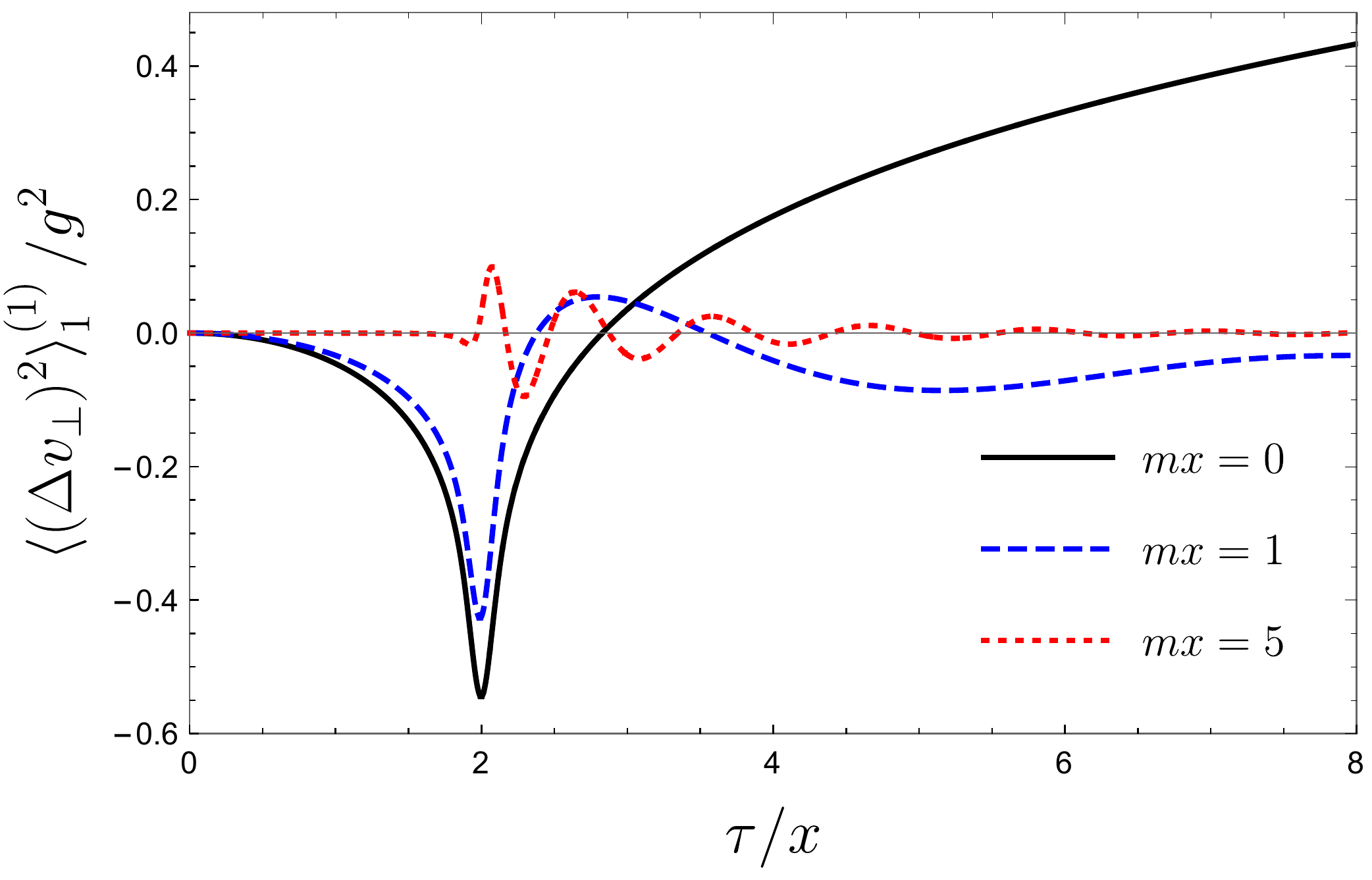}
\caption{Velocity dispersions $\braket{(\Delta v_{\bot})^2}^{(1)}_{1}$ as function of $\tau/x$ for some representative values of $mx$, where we set $n=20$. The solid curve recovers the regularization obtained in \cite{delorenci2014}. As $mx$ grows, the dashed and dotted curves show that the dispersions start to oscillate, a fingerprint of the non-Huygensian character of the background field coming from its mass.}
\label{fig4}
\end{figure}

The dispersions described by Eq.~\eqref{eq20} by setting $D=1$ are depicted in Fig.~\ref{fig4} for some illustrative values of $mx$.   
We see that the divergence at $\tau=2x$ is regularized, thus recovering the result in \cite{delorenci2014} for the massless case $mx=0$ via a completely different technique. From the dotted curve, for which $mx=5$, we see that the dispersion oscillates with a decreasing magnitude in a similar manner to the oscillations found in Sec.~\ref{secIII}.
\begin{figure*}
\subfloat[\label{fig5a}]{%
  \includegraphics[scale=0.55]{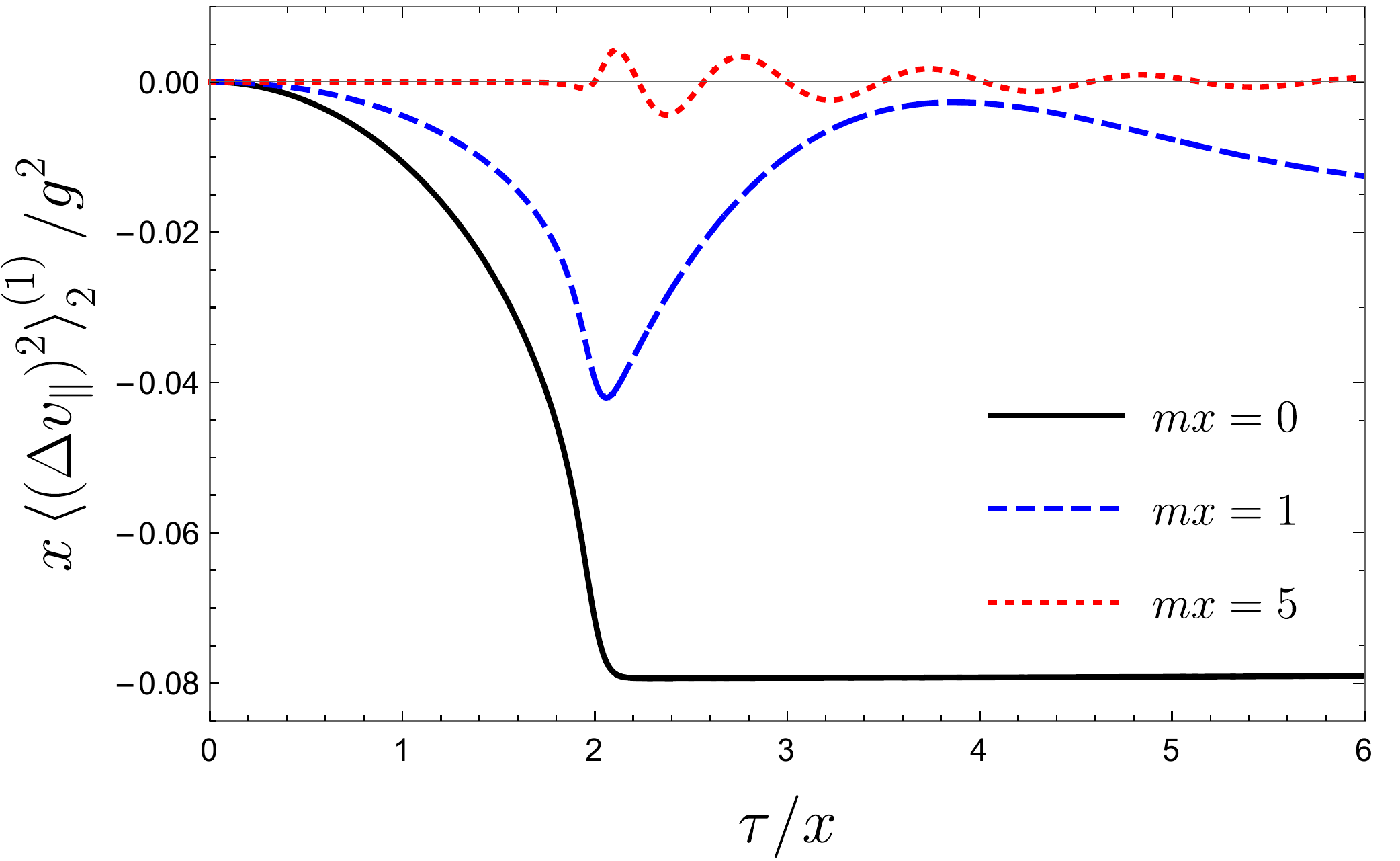}%
}
\\
\subfloat[\label{fig5b}]{%
  \includegraphics[scale=0.55]{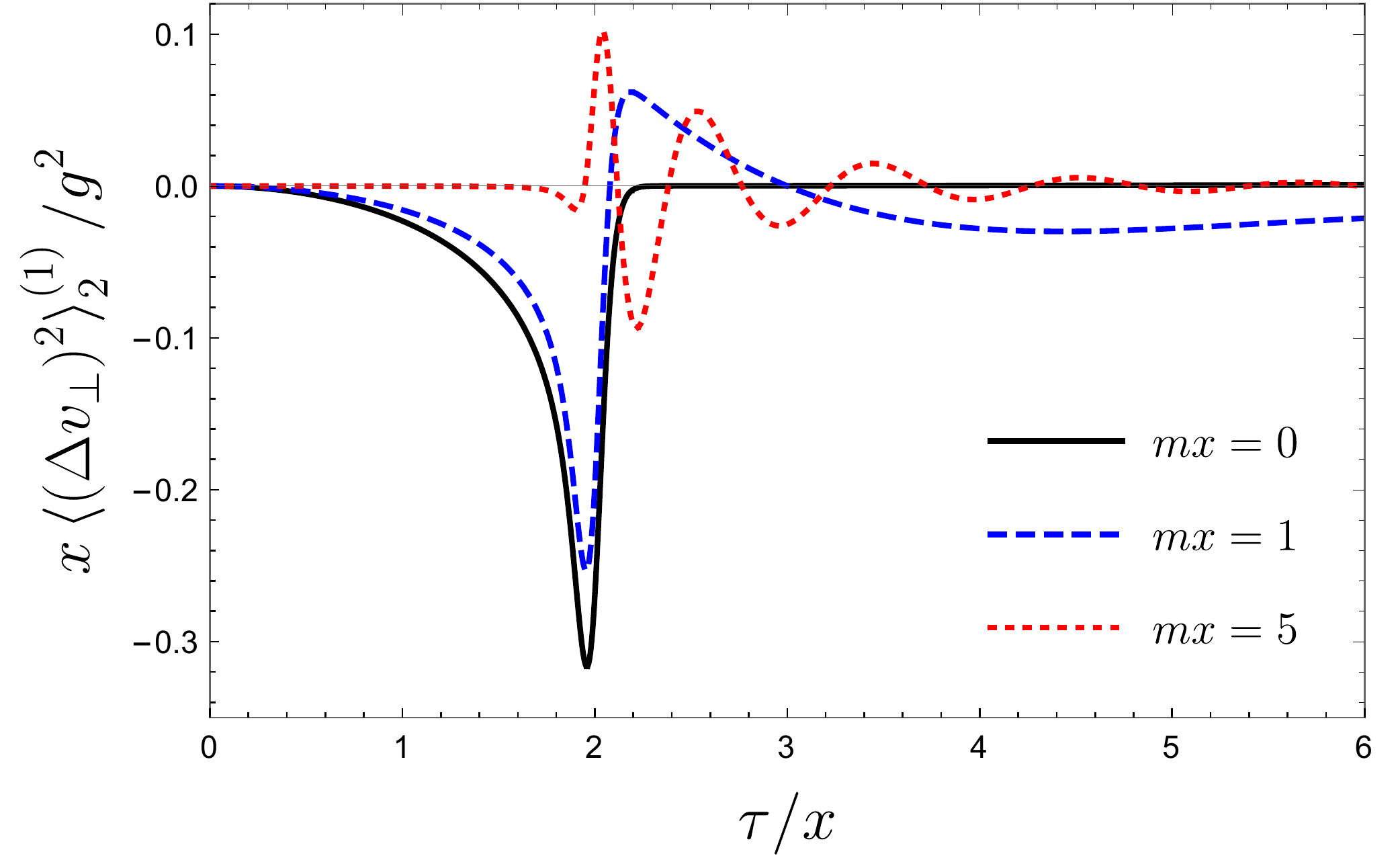}%
}
\caption{Velocity dispersions modeled with the switching mechanism $F^{(1)}_{n,\tau}$ for $D=2$. Here, $n=20$. As $D=2$, the velocity has dispersions parallel to the wall (a), and perpendicular to the wall (b). Again, if $mx>0$, the oscillatory behavior is recovered.}
\label{fig5}
\end{figure*}

When $D=2$, the background field does not satisfy the Huygens' principle even if the field mass vanishes, as anticipated. Thus, in this system the dispersions should present a behavior different from the $3+1$ and $1+1$ massless cases. In fact, the solid curves in Figs.~\ref{fig5a} and \ref{fig5b} show that both dispersions evolve as $\tau$ increases, and stabilize quickly just after $\tau=2x$, the parallel component approaching a negative constant, and the perpendicular one approaching zero. This should be compared with the solid curve in Fig.~\ref{fig4}, and with the results for the massless $3+1$ case \cite{Camargo2018}, where the dispersions do not present this sudden saturation. Another intriguing effect is the vanishing of the perpendicular velocity dispersion (solid curve in Fig.~\ref{fig5b}), when we compare with the electric charge, and the scalar charge in $3+1$ dimensions, where there is always a residual effect in the late time regime. As $mx$ increases, the same characteristic oscillatory behavior takes place in both directions, as expected.

Before passing to the analysis of the late time regime, we quote a couple of properties revealed by Eqs.~\eqref{eq19} and \eqref{eq20} for $D=3$. The limit $m\rightarrow0$ reproduces the results presented in Ref.~\cite{Camargo2018}, where the usual divergences at $x=0$ and $\tau=2x$ are regularized. As $mx$ increases, a transition to the oscillatory behavior occurs, and the dispersions have the same form as the ones in Fig.~\ref{fig5}.

\begin{figure}[h!]
\center
\includegraphics[scale=0.55]{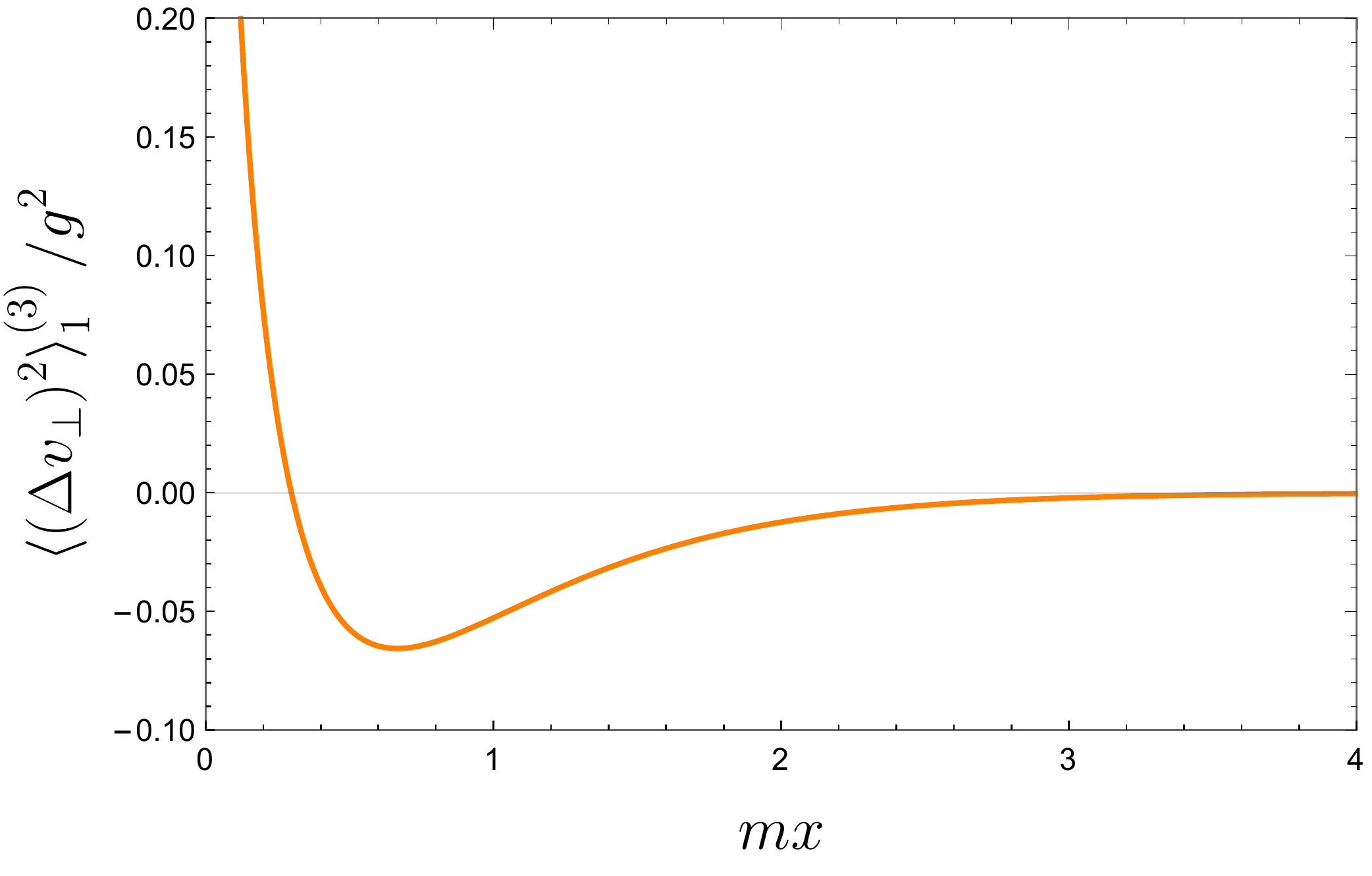}
\caption{Late time behavior of the velocity dispersion in $1+1$ modeled with the switching mechanism $F^{(3)}_{\tau_s,\tau}$ as a function of $mx$. As $mx\rightarrow0$, we recover the expected late time divergence of this model. For finite $mx$, the theory is regularized, showing that the divergence comes from the infrared sector of the theory.}
\label{fig6}
\end{figure}  

As discussed in Sec.~\ref{secIV}, in order to gain control of the late time regime, a suitable switching function must be implemented. A closed analytic expression can be obtained by using $F^{(3)}_{\tau_s,\tau}(t)$. From the definition given in Sec.~\ref{secIV}, and the result from Appendix \ref{Lemma}, the dispersions in the late time regime for any dimension $D$, mass $m$, and switching time $\tau_s$ are given by
\begin{align}
\braket{(\Delta v_{\|})^2}^{(3)}_{D}&=-\frac{g^2}{\pi x^{D-1}}\left[\frac{mx}{4\pi \sqrt{1+(\tau_s/x)^2}}\right]^{\frac{D-1}{2}}K_{\frac{D-1}{2}}\left[2mx\sqrt{1+(\tau_s/x)^2}\right],\label{eq21}\\
\braket{(\Delta v_{\bot})^2}^{(3)}_{D}&=8\pi x^2\braket{(\Delta v_{\|})^2}^{(3)}_{D+2}-\braket{(\Delta v_{\|})^2}^{(3)}_{D}.\label{eq22}
\end{align}
These equations have a plethora of possible applications, ranging from the determination of the energy exchanged due to the switching process to the infrared regularization in the $1+1$ case. Here, we explore some of them, related to already cited phenomena. Setting $D=3$ in Eqs.~\eqref{eq21} and \eqref{eq22}, and taking the limit $m\rightarrow0$,
\begin{align}
\braket{(\Delta v_{\|})^2}^{(3)}_{3}&\stackrel{m\rightarrow0}{=}-\frac{g^2}{8\pi^2}\frac{1}{x^2+\tau_s^2},\label{eq23}\\
\braket{(\Delta v_{\bot})^2}^{(3)}_{3}&\stackrel{m\rightarrow0}{=}-\frac{g^2}{8\pi^2}\frac{x^2-\tau_s^2}{(x^2+\tau_s^2)^2}.\label{eq24}
\end{align}
In the sudden switching limit, $\tau_s=0$, $\braket{(\Delta v_{\bot})^2}^{(3)}_{3}=\braket{(\Delta v_{\|})^2}^{(3)}_{3}=-g^2/8\pi^2x^2$, which is the reported residual effect in the $3+1$ massless case \cite{Camargo2018}. Equations \eqref{eq23}, \eqref{eq24} are physically equivalent to the ones obtained from the switching function $F^{(2)}_{\tau_s,\tau}(t)$ in Ref.~\cite{Camargo2018}, but have the advantage of being of easier manipulation.

Let us return to the one dimensional system, and set $D=1$ in Eq.~\eqref{eq22} to obtain
\begin{equation}
\braket{(\Delta v_{\bot})^2}^{(3)}_{1}=\frac{g^2}{\pi}\left\{K_{0}\left[2mx\sqrt{1+(\tau_s/x)^2}\right]-\frac{2mx}{\sqrt{1+(\tau_s/x)^2}}K_{1}\left[2mx\sqrt{1+(\tau_s/x)^2}\right]\right\}.\label{eq25}
\end{equation}
In the sudden switching regime, $\braket{(\Delta v_{\bot})^2}^{(3)}_{1}=(g^2/\pi)\left[K_{0}(2mx)-2mxK_{1}(2mx)\right]$, and thus the field mass acts as an infrared regulator for the reported divergence appearing in the late time regime of the massless case. As depicted in Fig.~\ref{fig6}, the dispersion positively diverges as $mx\rightarrow0$. For higher values of $mx$, it vanishes at $mx\approx0.298$, becomes negative, and then goes to zero as $mx$ goes to infinity. These results show that the late time regime for the $1+1$ dimensional case is highly dependent upon the mass of the scalar field.

We close this section with some remarks concerning the applicability of our formulas. Equations \eqref{eq19}, \eqref{eq20}, \eqref{eq21}, and \eqref{eq22} can be combined to describe velocity dispersions due to modified vacuum fluctuations in any dimension, for any field mass.
\begin{figure}[h!]
\center
\includegraphics[scale=0.55]{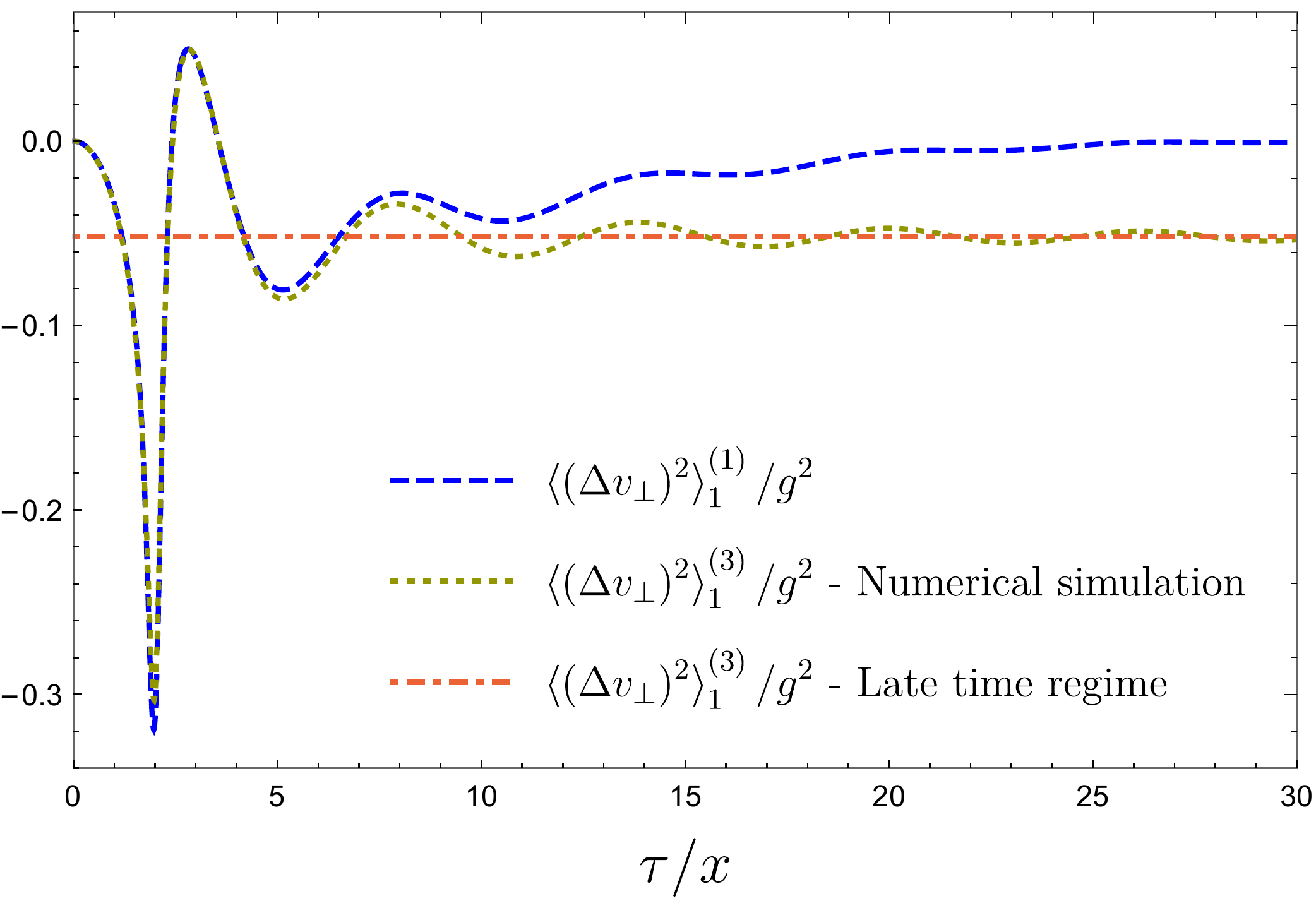}
\caption{Comparison of both switching process $F^{(3)}_{\tau_s,\tau}$ and $F^{(1)}_{n,\tau}$ for $D=1$ and $mx=1$. Here, $n=10$ and $\tau_s/x=0.1$. For small $\tau/x$, both process are indistinguishable, as shown by the dashed and dotted curves. As the late time regime is approached, the dispersion modelled by $F^{(3)}_{\tau_s,\tau}$ oscillates about its late time regime value, given by the dot-dashed curve.}
\label{fig7}
\end{figure}  
However, care must be taken depending on which regime we are working. For instance, as discussed in the previous section, the switching mechanism $F^{(1)}_{n,\tau}$ is such that the switching time $\tau_s$ is a linear function of the measuring time $\tau$. Thus, for small $\tau$, the system approaches the sudden switched one, and the near-boundary behavior cannot be studied. Despite this, the near-boundary behavior can be studied via the switching $F^{(3)}_{\tau_s,\tau}$ in the late time regime. Moreover, another feature coming from the use of switching functions is related to numerical simulations. For instance, the switching $F^{(3)}_{\tau_s,\tau}$ is such that the integral kernel in Eq.~\eqref{eq12} decays exponentially, and thus the integral converges rapidly. In this way, the near-boundary regime can be considered for intermediate times via numerical integration. Figure \ref{fig7} shows an example of how the late time regime is approached, and how the dispersions in the $1+1$ case calculated with both switchings starts to deviate from each other at intermediate times. 

\section{Final remarks}
\label{secVI}
In the previous sections we studied how a particular transition between scalar field vacuum states causes velocity fluctuations on a test particle interacting with this field. The transition experienced by the particle between the empty space and the space with a perfectly reflecting plane boundary was modeled by an analytic function acting as a sort of smooth switching mechanism. This method, with various applications explored in the literature, is based on the fact that the switching functions describing  smooth transitions regularize reported divergences by providing a certain scale for cutting-off nonphysical high frequency field modes, depending on the experimental arrangement in which the measurement is done. 
Here, we have explored further aspects of such systems by including scalar fields that can be found in nature, e.g., the massive Higgs field. The presence of the a massive field gives rise to characteristic oscillations in the dispersions of the velocity of a test particle. The main features of our work can be synthesized as follows. We studied how vacuum transitions induce velocity fluctuations of scalar particles. The quantization of the massive scalar field in Minkowski spacetime was revisited, stressing how nonlocal effects are unusual if the background field does not satisfies the Huygens' principle, and in particular, how velocity fluctuations were expected to be different for such background fields. In order to obtain analytic closed expressions for the dispersions, a systematic definition of realistic switching functions was presented by working in frequency space. This analysis revealed the mechanism behind the regularization process, which were applied to obtain the velocity dispersions in arbitrary dimensions, and for any field mass.
We showed, in particular, how they recover previous results, and solve open problems, as for instance the divergency at late times in the massless $D=1$ case, which comes from the infrared sector of the theory. 

It is noteworthy that in an actual experimentation of this effect, a real boundary would be present instead of the idealized one, and we do not expect to find any divergent outcome, as such boundary have to be transparent for very high frequency modes. In this way, renormalization with respect to Minkowski [Eq.~\eqref{eq11}] removes all the undesirable high frequency modes, rendering completely finite results for a sudden switched experiment. Nevertheless, a finite switching time is always present in such measurements, and can be important even if realistic boundaries are implemented. In fact, if the boundary natural frequency cutoff is higher than the one provided by experimentation, then we would obtain results similar to the ones presented in the previous sections for the idealized boundary. This shows that more than simply providing a way of finding finite results, the use of switching mechanisms is required by experimentation, and must be included in the calculation even for realistic boundaries. 

We close this work mentioning another important aspect presented by the previous analysis, that concerns subvacuum phenomena. It should have been noticed that the velocity dispersions depicted in Figs.~\ref{fig4}, \ref{fig5}, \ref{fig6}, and \ref{fig7} become negative in some range of its parameters. This is a truly quantum feature, as at classical level no ``negative dispersions'' can be found. Care must be taken, though, when interpreting such effects. In our case, these negative velocity dispersions simply mean that the effect of the boundary is to reduce by a certain amount the value of the dispersions coming from a more general model (considering for instance backreaction \cite{johnson2002}). Thus, in a real measurement one would obtain positive quantities lessen by subvacuum fluctuations.  

\acknowledgments
This work was partially supported by the Brazilian research agencies CNPq (Conselho Nacional de Desenvolvimento Cient\'{\i}fico e Tecnol\'ogico) under Grant No. 302248/2015-3, FAPEMIG (Funda\c{c}\~ao de Amparo \`a Pesquisa do Estado de Minas Gerais),  FAPESP (Funda\c{c}\~ao de Amparo \`a Pesquisa do Estado de S\~ao Paulo) under Grant No. 2015/26438-8, and CAPES (Coordena\c{c}\~ao de Aperfei\c{c}oamento de Pessoal de N\'{\i}vel Superior).

\appendix
\section{An expression for the late time behavior in the massive case (Modified Riemann-Lebesgue Lemma)}
\label{Lemma}
Let $f$ be in the space of absolute integrable functions in $(0,\infty)$, $L^{1}$. Then
\begin{equation}
\lim_{\tau\rightarrow\infty}\int_{0}^{\infty}\d k \frac{k\e^{i \tau\sqrt{k^2+m^2}}}{\sqrt{k^2+m^2}}f(k)=0,
\label{RLlemma}
\end{equation}
where $m\geq0$ is a parameter. The case $m=0$ is a particular case of the Riemann-Lebesgue lemma \cite{bochner}. As we have seen, the general expression for the dispersions can be obtained by
\begin{equation}
\braket{(\Delta v_i)^2}=g^2\left[\frac{\partial}{\partial x_i}\frac{\partial}{\partial x_i'}A(\x,\x';\tau)\right]_{\x'=\x},
\end{equation}
where
\begin{equation}
A(\x,\x';\tau)=
-\frac{1}{(2\pi)^{\frac{D}{2}}|\hat{\Delta}\x|^{\frac{D}{2}-1}}\int_0^\infty \d k \frac{k}{\omega}\left[1-\cos\left(\omega\tau\right)\right]\frac{\left|\mathcal{D}(\omega)\right|^{2}}{\omega^{2}}k^{\frac{D}{2}-1}J_{\frac{D}{2}-1}(k|\hat{\Delta}\x|),
\end{equation} 
$\omega=\sqrt{k^2+m^2}$. Notice that if $\left|\mathcal{D}(\omega)\right|^{2}$ is at least bounded as $k\rightarrow \infty$, then Eq.~\eqref{RLlemma} gives us the late time limit
\begin{equation}
A(\x,\x';\infty)=-\frac{1}{(2\pi)^{\frac{D}{2}}|\hat{\Delta}\x|^{\frac{D}{2}-1}}\int_0^\infty \d k \frac{k}{\omega^3}\left|\mathcal{D}(\omega)\right|^{2}k^{\frac{D}{2}-1}J_{\frac{D}{2}-1}(k|\hat{\Delta}\x|).
\end{equation}
A suitable choice for $\left|\mathcal{D}(\omega)\right|^{2}$ (as discussed in Sec.~\ref{secIV}) is
\begin{equation}
\left|\mathcal{D}(\omega)\right|^{2}=(1+2\tau_s\omega)\e^{-2\tau_s\omega}.
\end{equation} 
A closed form for $A(\x,\x';\infty)$ can then be obtained by using the results in \cite{gradshteyn}
\begin{equation}
A(\x,\x';\infty)=-\frac{2m^{\frac{D-3}{2}}}{(2\pi)^{\frac{D+1}{2}}\sqrt{4\tau_{s}^{2}+|\hat{\Delta}\x|^2}^{\frac{D-3}{2}}}K_{\frac{D-3}{2}}\left(m\sqrt{4\tau_{s}^{2}+|\hat{\Delta}\x|^2}\right).
\end{equation}
In order to prove Eq.~\eqref{RLlemma}, we may proceed as follows. Suppose first that $f(k)=\chi(k;a,b)\equiv\theta(k-a)\theta(b-k)$, $0\leq a<b<\infty$. This kind of function is called characteristic function. Then
\begin{equation}
\int_{0}^{\infty} \d k \frac{k \e^{i\tau\sqrt{k^2+m^2}}}{\sqrt{k^2+m^2}}f(k)=\int_{a}^{b} \d k \frac{k\e^{i\tau\sqrt{k^2+m^2}}}{\sqrt{k^2+m^2}}=\frac{\e^{i\tau\sqrt{b^2+m^2}}-\e^{i\tau\sqrt{a^2+m^2}}}{i\tau},
\end{equation}
and thus the result holds. By linearity, it also holds for any finite combination of characteristic functions
\begin{equation}
g(k)=\sum_{n=0}^{N}g_{n}\chi(k;a_{n},b_{n})
\label{simplefunction}
\end{equation}
with $0\leq a_{n}<b_{n}<\infty$ for all $n$, $g_{n}$ being constant complex numbers. Now let $f \in L^{1}$ be arbitrary. We will use the fact that the set of functions like those given by Eq.~\eqref{simplefunction} is dense in $L^{1}$ \cite{bochner}. This means that given $\epsilon>0$, there exists a function of the form $g(k)$ for some $N$ such that
\begin{equation}
\int \d k|f(k)-g(k)|<\frac{\epsilon}{2},
\end{equation}
and we will take $\tau$ sufficiently large to ensure
\begin{equation}
\left|\int \d k \frac{k\e^{i\tau\sqrt{k^2+m^2}}}{\sqrt{k^2+m^2}}g(k)\right|< \frac{\epsilon}{2}.
\end{equation}
Finally, we can write
\begin{equation}
\int \d k \frac{k\e^{i\tau\sqrt{k^2+m^2}}}{\sqrt{k^2+m^2}}f(k)=\int \d k \frac{k\e^{i\tau\sqrt{k^2+m^2}}}{\sqrt{k^2+m^2}}[f(k)-g(k)]+\int \d k \frac{k\e^{i\tau\sqrt{k^2+m^2}}}{\sqrt{k^2+m^2}}g(k)
\end{equation}
and by noticing that $k<\sqrt{k^2+m^2}$,
\begin{align}
\left|\int \d k \frac{k\e^{i\tau\sqrt{k^2+m^2}}}{\sqrt{k^2+m^2}}f(k)\right|&\leq \int \d k \left|f(k)-g(k)\right|+\left|\int \d k \frac{k\e^{i\tau\sqrt{k^2+m^2}}}{\sqrt{k^2+m^2}}g(k)\right|\nonumber\\
&<\frac{\epsilon}{2}+\frac{\epsilon}{2}=\epsilon.
\end{align}

\end{document}